\documentclass[12pt]{article}
\usepackage{amsfonts,amsthm,amsmath,amssymb,upgreek,mathrsfs,latexsym,subcaption,physics,float,bm}
\usepackage{hyperref}
\usepackage{cite}
\usepackage[paper=letterpaper,margin=1in]{geometry}
\usepackage{graphicx}
\usepackage{units}

\begin{document}
\thispagestyle{empty}
\renewcommand{\thefootnote}{\fnsymbol{footnote}}

\begin{center}

\bf{{\LARGE Notes on the Squashed Sphere Lowest Landau Level}\\
\begin{center}
   \bf {Jeff Murugan, Jonathan P. Shock \& Ruach Pillay Slayen }\\
   \bigskip 
   \rm
   \bigskip
   {\it Laboratory for Quantum Gravity \& Strings,\\ 
   Department of Mathematics and Applied Mathematics,\\ 
   University of Cape Town, South Africa}\\
   \rm
\end{center}
\vspace{1cm}}

{\bf Abstract}
\end{center}
\begin{quotation}
\noindent
In a recent article \cite{Murugan:2018hsd}, we were motivated by the question of whether any of the remarkable condensed matter phenomena, such as the quantum Hall effect (QHE), the Integer quantum Hall effect (IQHE) {\it etc.}, could potentially be observed in the extreme astrophysical environments of neutron stars.  As a prequel to that work, and with the aim of understanding better the role of the {\it geometry} of the conducting surface on the structure of Landau levels, in this article we study the quantum dynamics of a quantum particle on a squashed sphere. More specifically, we study the dynamics of a single particle on an oblate squashed Haldane sphere {\it i.e.} a 2-sphere enclosing a single magnetic monopole at its 
        center. While several features of the conventional Haldane sphere persist, by numerically solving the Schr\"odinger equation in this background, we find that the particle becomes increasingly localised in a band between the equator and the poles, with a corresponding increase of the eccentricity of the spheroid.
\end{quotation}

\setcounter{page}{0}
\setcounter{tocdepth}{2}
\setcounter{footnote}{0}
\newpage
 \tableofcontents
\section{Introduction}\label{intro}
Ever since Haldane's seminal work \cite{Haldane:1983xm} introducing the spherical geometry enclosing a magnetic monopole as a device to circumvent the subtleties associated with boundaries, quantum Hall states have been extensively studied on a variety of compact and non-compact, curved and flat two-dimensional manifolds. In some sense this culminated Dunne's synthesis \cite{Dunne:1991cs} of the  spherical, planar and hyperbolic geometries into a common framework. This in turn laid the groundwork for the treatment of the IQHE \cite{Klevtsov:2013iua} and fractional quantum Hall effect (FQHE) \cite{Can:2014awa} states on surfaces of arbitrary curvature. Particular studies in the recent past have included the quantum Hall systems on tori \cite{tor}, cylinders \cite{cyl} and higher genus Riemann surfaces \cite{rie}. Although largely theoretical at present, such studies have greatly enriched our understanding of experimentally observable states of quantum matter. Indeed, studies of ground states on curved surfaces were found to provide a complete description of the QHE on a flat background,  and to uncover universal features of the QHE inaccessible to calculations in flat space \cite{Can:2014awa}. All dissipation-free transport coefficients of the QHE at low energies were understood as the response of the ground state to changes in scalar curvature on a closed manifold \cite{Can:2014awa}. Cumulatively, these results have highlighted the importance of {\it geometry} - as opposed to {\it topology} -  in our understanding of the QHE \cite{Avron:1995fg,lev}. We would like to understand this connection between the QHE and the geometry of the conducting surface a little deeper. Toward this end, in this article we study quantum states of matter in a class of deformations of the Haldane sphere in which the round sphere surrounding the magnetic monopole is deformed into an oblate spheroid, or squashed sphere.\\

\noindent
Key to this problem is the fact that, apart from at the poles, the resulting magnetic field is nowhere perpendicular to the 2-surface. Consequently, the projecting normal to the sphere is an inhomogenous magnetic field. Surprisingly, comparatively little attention has been paid to 2-dimensional quantum states in inhomogeneous magnetic fields. In the context of open planar systems, most of the work carried out to date has focused either on small deviations from homogeneity, such as fields which approach a uniform value at large distances \cite{2d}, or on highly constrained variations, such as fields which are monotonic functions of radial distance \cite{2a}. In the context of compact surfaces, the systems studied generally assume small variations over a large constant background \cite{Can:2014awa}. Under the same assumptions, we will affect a deformation of the Haldane sphere and study the structure of the Landau levels of a quantum particle confined to an ellipsoidal surface parameterised by the eccentricity $e$. We squash the sphere in the presence of the fixed background magnetic field to produce a magnetic flux density that is inhomogeneous over the ellipsoid. The resulting system breaks the $SO(3)$ symmetry of the Haldane sphere to a $U(1)$, with consequent implications for the single-particle wanvefunctions.\\

\noindent
In Section 2 we review Dunne's unified framework \cite{Dunne:1991cs} to treat charged particles on 2-dimensional surfaces immersed in a constant (homogeneous) perpendicular magnetic field with the goal of generalising the formalism to treat the case of interest. This is followed by our treatment of the squashed sphere system in Section 3 where we show how to extend the treatment in \cite{Dunne:1991cs} to a particle confined to the deformed sphere in both uniform as well as non-uniform magnetic fields. We conclude in Section 4 with some contextual comments about the relevance of the geometrically deformed Haldane problem in high energy and theoretical condensed matter physics.

\section{A General Framework}
In this section, following the formalism developed in \cite{Can:2014awa,Dunne:1991cs}, we review the QHE in an arbitrary K\"ahler geometry and use it to compute the $g_{s}=2$ wavefunctions of charged particles confined to two-dimensional surfaces in a constant, perpendicular magnetic field. We will then show how this may be extended to non-constant magnetic flux density and $N$ free fermions using the Slater determinant. To begin, let's consider the single-particle case.

\subsection{Single-Particle States}

To simplify the discussion, we will work in isothermal complex coordinates $z=x_1+ix_2$, $\bar{z}=x_1-ix_2$ with holomorphic and anti-holormorphic derivatives defined as $\partial = \frac{1}{2}(\partial_1-i\partial_2)$ and $\bar{\partial} = \frac{1}{2}(\partial_1+i\partial_2)$ respectively. Our particle will be confined to a Riemann surface with line element $ds^2=g_{z\bar{z}}dzd\bar{z}\equiv\sqrt{g}dzd\bar{z}$. The volume form on the manifold,
\begin{equation}\label{vol-form}
    dV=\frac{\sqrt{g}}{2i} dz\wedge d\bar{z}\,,
\end{equation}
while its Ricci scalar curvature is given by
\begin{equation}
    \text{Ric} = -\Delta_g\log\sqrt{g}\,,
\end{equation}
 where the Laplace-Beltrami operator $\displaystyle \Delta_g \equiv \frac{4}{\sqrt{g}}\partial\bar{\partial}$.
In the cases of interest to us, the metric on the surface may be expressed in terms of the K\"ahler potential $K$, defined through
\begin{equation}
    \partial\bar{\partial}K=\sqrt{g}\,.
\end{equation}
To define the magnetic field in which the particle moves, we will need some notion of \textit{orthogonality} in the two-dimensional geometric framework in which we are working. To this end, the most natural definition of the constant, perpendicular gauge field is one whose 2-form field strength is proportional to the volume form $F = BdV$. The 2-form field strength can also be expressed in terms of the gauge potential
\begin{equation}
    F = (\partial \bar{A}-\bar{\partial}A)dz\wedge d\bar{z}\,,
\end{equation}
where $A\equiv A_z$ and $\bar{A}\equiv A_{\bar{z}}$ are the holomorphic and anti-holomorphic components of the gauge potential respectively. From this, it follows that
\begin{equation}\label{b}
   \partial \bar{A}-\bar{\partial}A= i\sqrt{g}B/2\,.
\end{equation}
Moreover,  in covariant Coulomb gauge, the gauge potential also satisfies $\bar{\partial}A+\partial\bar{A} = 0$, or equivalently, in terms of the magnetic field
\begin{equation}\label{Bnow}
   B=\frac{4i}{\sqrt{g}}\bar{\partial}A\,.
\end{equation}
If we define the real \textit{magnetic potential} $Q$ through the relations
\begin{align}\label{defQ}
i\hbar\partial Q = 2eA\,, \qquad
-i\hbar\bar{\partial} Q = 2e\bar{A}\,,
\end{align}
and combine (\ref{defQ}) and (\ref{Bnow}), we find that the magnetic potential satisfies the second order differential equation
\begin{equation}\label{Q}
    \Delta_g Q \equiv \frac{4}{\sqrt{g}}\partial\bar{\partial} Q = -\frac{2eB}{\hbar}\,.
\end{equation}
In the case of a constant magnetic flux density through the surface (constant $B$), it is easily verified that the magnetic potential $Q$ can be chosen to be proportional to the K\"ahler potential of the surface, or
\begin{equation}\label{QK}
    Q = -\frac{K}{2l^2}\,,
\end{equation}
where $l \equiv \sqrt{\hbar/eB}$ is the magnetic length.\\

\noindent
With all geometric objects defined, we now turn to the physics. The Pauli Hamiltonian for spin polarized electrons appropriate for modelling free electrons on a Riemann surface \cite{2} is
\begin{equation}
    H = \frac{1}{2m}\Big( \frac{1}{\sqrt{g}}\pi_i \sqrt{g}g^{ij}\pi_j -\frac{g_s}{2}e\hbar B \Big)\,,
\end{equation}
where $\pi_i = -i\hbar\partial_i-eA_i$ is the kinetic momentum with $i,j = 1,2$, and $g_s$ is the Land\`e $g$-factor. Changing to complex coordinates and using the commutation relations,
\begin{equation}
    [\bar{\pi},\pi] =\sqrt{g}\frac{e\hbar B}{2}\,,
\end{equation}
puts the Hamiltonian in the form
\begin{equation}\label{ham}
     H = \frac{2}{m}\Big( \frac{1}{\sqrt{g}} \pi\bar{\pi} + \frac{2-g_s}{8}e\hbar B \Big)\,,
\end{equation}
where $\pi = -i\hbar\partial - eA$ and $\bar{\pi} = -i\hbar\bar{\partial} - e\bar{A}$ are the holomorphic and anti-holomorphic components of the momentum respectively. 

\subsubsection{Uniform Magnetic Flux Density}

As a check of the formalism, let's now make contact with the descriptions of the familiar planar and spherical monopole systems. Setting $\hbar=m=e=1$ and focussing on the case of a uniform magnetic flux, where $B$ is constant and (\ref{QK}) holds, the Hamiltonian reads
\begin{align}\label{hamgen}
    H = -\frac{2}{\partial\bar{\partial}K} \Big(\partial-\frac{1}{4l^2}\partial K\Big)\Big(\bar{\partial}+\frac{1}{4l^2}\bar{\partial} K\Big) + \frac{2-g_s}{4}B\,.
\end{align}
For $g_s=0$, this is precisely the form of the Hamiltonian for the planar and spherical monopole systems, modulo a rescaling of coordinates by $\sqrt{2/B}$ and once the respective K\"ahler potentials for the plane and the sphere have been substituted in
\begin{equation}\label{Ksphere}
    K=\begin{cases} 
      |z|^2 & \quad \text{plane} \\
      4R^2\log(1+|z|^2/4R^2) & \quad \text{sphere of radius $R$.}  
      \end{cases}
\end{equation}
Given the form of the Hamiltonian (\ref{hamgen}), it is natural to redefine the wavefunctions as
\begin{align}\label{redef3}
    \psi(z,\bar{z})=e^{-K/4l^2}\hat{\psi}(z,\bar{z})\,,
\end{align}
where the $\hat{\psi}$'s are Hilbert space elements, with inner product
\begin{align}\label{prodgen}
\bra{f}\ket{g} = \mathcal{N} \int dV e^{-K/2l^2} \overline{f(z,\bar{z})}g(z,\bar{z})\,,
\end{align}
and normalization constant $\mathcal{N}$.  The Hamiltonian  corresponding to (\ref{hamgen}) which acts on $\hat{\psi}$ is then given by
\begin{align}\label{redhamgen}
    \hat{H} &= e^{K/4l^2}He^{-K/4l^2} \nonumber \\
    &= -\frac{2}{\partial\bar{\partial}K} \Big(\partial-\frac{1}{2l^2}\partial K\Big)\bar{\partial} + \frac{2-g_s}{4}B\,.
\end{align}
Clearly any holomorphic function will be annihilated by the first term in this Hamiltonian. For $g_s=0$, such states will all be degenerate with energy $B/2$, and will constitute the lowest Landau level (LLL) of the free fermion. The full set of LLL states in this case will be given by
\begin{align}
    \psi_{m}(z,\bar{z}) = s_m(z)e^{-K/4l^2}\,,
\end{align}
where the holomorphic functions $s_m$ satisfy $\bar{\partial}s_m=0$ such that $\psi_m$ is normalisable under the inner product
\begin{equation}
    \langle\psi_n|\psi_m \rangle \equiv \int s_n \bar{s}_m e^{-K/2l^2}dV = \delta_{mn}\,.
\end{equation}
Our treatment here has been general\footnote{In fact it is slightly more general even. Since all that is required in this formalism is that the {\it normal component} to the conducting surface is constant, it is also applicable in the setting where neither the magnetic field itself nor the normal vector field to the surface are constant but their dot product is.} and applies to any system of charged particles confined to a 2-dimensional K\"ahler Riemann surface in a constant and perpendicular magnetic field. Notice that the form of the exponential measure factor of the LLL states is entirely determined by the underlying geometry of the Riemann surface, via the K\"ahler potential. More precisely, the $s_n$ are sections of a holomorphic line bundle equipped with hermitian metric $e^{-K/2l^2}$. Holomorphic sections defined in the conformal class of a sphere are polynomials, whose degree cannot exceed the number of flux quanta piercing the 2-surface \cite{Can:2014awa}
\begin{equation}
N_{\phi}= \frac{\Phi}{\Phi_0}\,,
\end{equation}
where $\Phi$ is the net flux through the surface and $\Phi_0=2\pi\hbar/e$ is known as the \textit{flux quantum}. It then follows from the Riemann-Roch theorem \cite{2} that the number of such holomorphic sections on a manifold of genus $G$ is given by
\begin{align}
    N=N_{\phi}-G+1\,.
\end{align}
Setting $G=0$ in this formula gives the correct counting for the degeneracy of the LLL in the spherical monopole system. Any $N_{\phi}+1$ linearly independent holomorphic polynomials of degree less that $N_{\phi}$ will thus furnish a basis for the LLL. The simplest such choice is
\begin{align}
    s_m(z)=z^m \quad \text{where} \quad m=0,1,...,N_{\phi}\,,
\end{align}
which coincides with the usual angular momentum eigenstates encountered on the sphere. Any K\"ahler Riemann surface which  possesses an additional azimuthal rotational symmetry will have an associated K\"ahler potential that depends only on $|z|^2$. In this case, the angular momentum operator $J=z\partial-\bar{z}\bar{\partial}$ commutes with the reduced Hamiltonian (\ref{redhamgen}). We may therefore look for simultaneous eigenstates of the form
\begin{align}
    \hat{\psi}=z^mP(|z|^2)\,.
\end{align}
Requiring that this state be an energy eigenstate leads to the following second order differential equation for $P$
\begin{align}
    \Big(\frac{x}{K'+xK''}\Big)P''+\frac{(m+1-x)K'}{(K'+xK'')}P'+\Big(\frac{E}{B}-\frac{1}{2}\Big)P=0\,,
\end{align}
where $x\equiv|z|^2$ and prime denotes a derivative with respect to $x$. For choices of $K$ for which this differential equation is solvable for $P$, what we have here is a prescription for finding the higher Landau level states. Indeed, substituting in for the K\"ahler potential for the plane yields the Laguerre equation of the planar system, while the  K\"ahler potential for the 2-sphere and redefining of the argument of $P$ so that
\begin{equation}
    P=P\Big( \frac{1-|z|^2/2BR^2}{1+|z|^2/2BR^2} \Big)\,,
\end{equation}
yields the Jacobi equation of the Haldane sphere\footnote{ See \cite{Dunne:1991cs} for the application of this framework to the \textit{hyperbolic} monopole system.}.

\subsubsection{Non-Uniform Magnetic Flux Density}

We now turn to a more general case in which the magnetic field through the surface is non-uniform ({\it i.e.} non-constant and/or non-perpendicular). We assume small variations over a large constant background \cite{Can:2014awa}. Due to the term proportional to $B=B(z,\bar{z})$ in the Hamiltonian (\ref{ham}), the degeneracy of the LLL is broken for general $g_s$. However, for $g_s=2$, the degeneracy persists and the LLL eigenstates states satisfy \cite{14}
\begin{equation}
    \bar{\pi}\psi = 0\,,
\end{equation}
as before. The solutions are now given by
\begin{equation}\label{soln}
    \psi_m(z,\bar{z}) = s_m(z)e^{\frac{1}{2}Q}\,,
\end{equation}
where $Q$ is the magnetic potential satifsying (\ref{Q}) and the holomorphic functions $s_m$ satisfy $\bar{\partial}s_m=0$ such that $\psi_m$ is normalisable with respect to the inner product
\begin{equation}\label{prodQ}
    \langle\psi_n|\psi_m \rangle \equiv \int s_n \bar{s}_m e^{Q}dV \,.
\end{equation}
The maximal degeneracy is still $N = N_{\phi}+1$ for an LLL basis of states $s_m(z)=z^m$ with $m=0,1,...,N_{\phi}$.

\subsection{Multi-Particle States}
Up until now we have only considered single particle states. To construct the $N_{p} $-particle ground state wavefunction for free fermions, we take the Slater determinant of the single particle states \cite{12},
\begin{equation}
    \Psi(\chi_1,...,\chi_{N_p}) = \frac{1}{\sqrt{N!}}e^{-\sum_i^{N_p}K(\chi_i)/4l^2}\det[s_m(z_i)]\,,
\end{equation}
where $N_p$ cannot exceed the maximal degeneracy $N$, and $\chi_i=(z_i,\bar{z}_i)$ is used to indicate that the argument is not holomorphic. For a fully filled Landau level, the Vandermonde identity implies that 
\begin{equation}
    \det[s_m(z_i)]= \mathcal{N}\sqrt{N!}\prod_{i<j}^N(z_i-z_j)\,,
\end{equation}
which is valid for spherical, conical and planar geometries. The maximally degenerate $N$-particle state is then given by
\begin{equation}
    \Psi = \mathcal{N}\prod_{i<j}^N(z_i-z_j)e^{-\frac{1}{4l^2}\sum_i^{N_p}K(\chi_i)}\,,
\end{equation}
where $\mathcal{N}$ is chosen such that $\bra{\Psi}\ket{\Psi}=1$. In the non-uniform magnetic flux case, the $N$-particle state is given by
\begin{align}
    \Psi = \mathcal{N}\prod_{i<j}^N(z_i-z_j)e^{\frac{1}{2}\sum_i Q(\chi_i)}\,.
\end{align}

\section{The Squashed Sphere}

Having laid out the formalism, let's now consider a variant of the Haldane sphere in which the sphere is squashed into an oblate spheroid. Specifically, the background monopole field will be left untouched while the sphere parametrically deformed from the round sphere into a spheroid. We begin by deriving the conformal map from the spheroid to the sphere, as required for the application of the general framework set out above.\\

\noindent
By stereographically projecting from the north pole onto the complex plane tangent to the south pole, the metric and K\"ahler potential on a round 2-sphere of radius $R$ can be written as
\begin{equation}
   \sqrt{g_0} = \frac{1}{(1+|z|^2/4R^2)^2}\,,
\end{equation}
and
\begin{equation}\label{K_0}
    K_0 = 4R^2 \log(1+|z|^2/4R^2)\,,
\end{equation}
respectively. Correspondingly the line element in polar coordinates is
\begin{equation}\label{sphere}
    ds_{sphere}^2 = \frac{1}{(1+r^2/4R^2)^2}(dr^2+r^2d\phi^2)\,.
\end{equation}
A manifold with metric $\sqrt{g}$ and K\"ahler potential $K$ is said to be conformally related to the metric $\sqrt{g_0}$ if $\sqrt{g}= e^{2\sigma}\sqrt{g_0}$ for some conformal factor $e^{2\sigma}$, which will in general be a function of the coordinates on the manifold. The K\"ahler potentials of the two metrics are related by 
\begin{equation}
    K = K_0 + u\,.
\end{equation}
where the \textit{deformed} part of the Kahler potential $u$ satisfies the Liouville equation
\begin{equation}\label{u}
    \Delta_{g_{0}} u \equiv \frac{4}{\sqrt{g_0}} \partial\bar{\partial}u = 2(e^{2\sigma}-1)\,.
\end{equation}
In the uniform field case, a basis for the LLL eigenstates on the surface with metric $\sqrt{g}$ is then given by
\begin{align}\label{ustates}
    \psi_m(z,\bar{z}) = z^m e^{-(K_0+u)/4l^2}, \quad m=0,1,...,N_{\phi}\,.
\end{align}
Note that in the limit $u\rightarrow 0$ these states reduce to the LLL states of the Haldane sphere, as expected. In the presence of a non-uniform field, the LLL eigenstates are given, as before, by
\begin{align}\label{Qstates}
    \psi_m(z,\bar{z}) = z^m e^{Q/2}, \quad m=0,1,...,N_{\phi}\,,
\end{align}
where $Q$ satisfies (\ref{Q}). We now seek to write down the LLL states on the surface of the oblate spheroid,  that we will refer to as a `squashed sphere'. This requires us to write down the conformal factor relating the metric of the spheroid to that of the sphere, and solve (\ref{u}) to find the K\"ahler potential of the spheroid. From this we can compute the LLL states via the  prescriptions given in \eqref{ustates} and \eqref{Qstates}.
\\

\subsubsection{Conformal Map from the Spheroid to the Sphere}

\begin{figure}
    \centering
	 \includegraphics[scale=0.4]{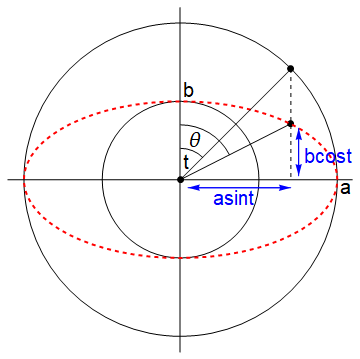} 
    \caption{Relationship between the parameter $\chi$ and the standard polar coordinate $\theta$}\label{tangle}
\end{figure}

We begin by constructing the conformal factor $e^{2\sigma}$. We know that the metric of the sphere $g^{S}_{ab}$ of radius $R$ and that of the plane $g^{P}_{ab}$ are related via stereographic projection as
\begin{equation}\label{ya}
g^{S}_{ab} (r,\phi) = \frac{1}{(1+r^2/4R^2)^2}g^{P}_{ab}(r,\phi)\,,
\end{equation}
where
\begin{equation}
g^{P}_{ab}(r,\phi) = 
\left(\begin{array}{cc} 
    1 & 0 \\ 
   0 & r^2 
\end{array}
\right)
\end{equation}
is the usual metric of the plane in polar coordinates and the stereographic projection is given by
\begin{align}
    r(\theta) &= 2R\cot(\theta/2)\,,
\end{align}
where $0<\theta<\pi$ is the polar angle on the sphere. To find the conformal factor relating the metric of the spheroid $g^{E}_{ab}$ to that of the plane, notice that the spheroid is defined in three-dimensional Cartesian coordinates by
\begin{equation}
    \frac{x^2+y^2}{a^2} + \frac{z^2}{b^2} = 1\,,
\end{equation}
where $a$ and $b$ are the equatorial and polar radii respectively. Alternatively, it can be written in parametric form as
\begin{align}\label{ellipsoidcoords}
    x&=a\sin\chi\cos\phi\,,  \nonumber \\
    y&=a\sin\chi\sin\phi\,,  \nonumber \\
    z&=b\cos\chi\,,
\end{align}
where $0\leq\chi\leq\pi$ and $0<\phi\leq2\pi$. While $\phi$ is the usual azimuthal angle, the parameter $\chi$ is not equal to the polar angle $\theta$ that the point $(x,y,z)$ makes with the $xy$-plane. It is referred to as the \textit{eccentric anomaly} in astronomy and has a geometric meaning illustrated in Figure \ref{tangle}. The relationship between $\chi$ and the standard polar angle $\theta$ is given by
\begin{equation}\label{hire}
\chi(\theta)= \text{arccot}\Bigg[ \frac{b}{a} \cot \Big[\theta-\frac{\pi}{2}\Big] \Bigg] + \frac{\pi}{2}\,.
\end{equation}
Note that the values of $\chi$ and $\theta$ coincide at the equator and the poles for any $a$ and $b$. In the spherical limit $a=b$, we of course recover $\chi(\theta)=\theta$ for all $\theta$.
 In these coordinates, the spheroidal line element is given by
\begin{equation}\label{ds'}
    ds_{E}^2 = (a^2\cos^2 \chi + b^2\sin^2\chi )d\chi^2 + a^2\sin^2 \chi d\phi^2\,.
\end{equation}
Next, consider the coordinate transformation\footnote{Note that, as explained in Appendix B, the map (\ref{r}) is constructed using a map to the sphere of radius $R$ as an intermediary step. This has introduced into the map the free parameter $R$, which is not fixed by the geometry of the spheroid. In our case we are subsequently mapping back to the sphere, so we of course identify this free parameter with the radius of this sphere.} \cite{3},
\begin{equation}\label{r}
    r(\chi) = 2 R e^{-h(\chi)}\,,
\end{equation}
where 
\begin{equation}\label{g}
    h(\chi) = \int^{\chi}_{\pi/2}\sqrt{\Big(\frac{b}{a}\Big)^2 + \cot^2v}\,dv\,.
\end{equation}
It follows that $r(\chi)$ satisfies the differential equation
\begin{eqnarray*}
    \frac{dr}{d\chi}&=-2R e^{-h} \frac{dh}{d\chi}
\end{eqnarray*}
or, equivalently, $dr^2 =4R^2 e^{-2h} \big[(b/a)^2+\cot^2 \chi \big]d\chi^2$, allowing us to rewrite the spheroidal line element (\ref{ds'}) as
\begin{equation}
    ds_{E}^2= \Big(\frac{a}{r}\sin \chi\Big)^2 \big(dr^2 +r^2d\phi^2\big)\,.
\end{equation}
In other words, the coordinate transformation (\ref{r}) conformally maps the spheroid of equatorial radius $a$ and polar radius $b$ to the plane. Consequently, the metric of the spheroid $g^{E}_{ab}$ and that of the plane $g^{P}_{ab}$ are related by
\begin{equation}\label{met}
   g^{E}_{ab} (r,\phi) = \Big( \frac{a}{r}\sin\chi \Big)^2 g^{P}_{ab} (r,\phi)\,.
\end{equation}
As a check, note that in the special case when $a=b=R$ (where $\chi=\theta$),
\begin{align}
    h(\chi)\rvert_{a=b=R} &= \int^{\theta}_{\pi/2}\sqrt{1 + \cot^2v}dv = \int^{\theta}_{\pi/2} \frac{1}{\sin v}dv = -\log\Big(\frac{\sin\theta}{1-\cos\theta}\Big)\nonumber\\
    \Rightarrow r\rvert_{a=b=R} &= 2Re^{-h} = 2R\frac{\sin\theta}{1-\cos\theta} = 2R\cot(\theta/2) \,. \label{reps0}
\end{align}
and this metric transformation from the spheroid to the plane reduces to the transformation from the sphere to the plane. We can now combine (\ref{met}) with (\ref{ya}) to recover the conformal relation between the metrics of the sphere and the spheroid:
\begin{equation}
    g^{E}_{ab}(r,\phi) = \Big( \frac{a}{r} \sin\chi \Big)^2 \Big(1+\frac{r^2}{4R^2}\Big)^2 g^{S}_{ab}(r,\phi)\,.
\end{equation}
Finally, changing coordinates from $r$ to $\chi$ gives the conformal factor relating the sphere and the spheroid,
\begin{equation}\label{conf}
    e^{2\sigma} = \Big( \frac{a}{2R}e^{h(\chi)} \sin\chi \Big)^2 \Big(1+e^{-h(\chi)}\Big)^2\,.
\end{equation}
In the case where $a=b=R$, this conformal factor reduces to unity, as expected. Note that in terms of these coordinates, $z=2Re^{-h(\chi)}e^{i\phi}$ and $\bar{z}=2Re^{-h(\chi)}e^{-i\phi}$ so that the Jacobian is given by $dz d\bar{z} = 8iR^2e^{-2h(\chi)}\sqrt{(b/a)^2 + \cot^2 \chi}d\phi\, d\chi$ and the metric becomes
\begin{align}\label{Qmetric}
   \sqrt{g} = \frac{1}{(1+r^2/4R^2)^2} = \frac{1}{(1+e^{-2h(\chi)})^2}\,,
\end{align}
from which we can construct the volume element 
\begin{align}\label{volQ}
 dV = 4R^2\frac{e^{-2h(\chi)}}{(1+e^{-2h(\chi)})^2}\sqrt{(b/a)^2 + \cot^2\chi}d\phi d\chi\,.
\end{align}
Note that in the limit $a=b=R$ our volume element reduces to the volume element of the sphere $dV = R^2\sin\theta\, d\phi\, d\theta$ as required.
\\

We want to squash the sphere in a volume preserving way. To this end it is useful to reformulate the above in terms of the eccentricity $e$ of the ellipsoid, defined by
\begin{equation}\label{ecc}
\frac{b^2}{a^2}=1-e^2, \quad 0<e<1 \,.
\end{equation}
Clearly as $e\rightarrow0$ we recover the spherical geometry, while $e\rightarrow1$ corresponds to the limiting case of an infinitely elongated oblate spheroid. The volume of our spheroid is given by 
\begin{equation}
V = \frac{4}{3}\pi a^2 b \,.
\end{equation}
Fixing $V=4\pi/3$, we recover the condition for volume preservation
\begin{equation}
b=\frac{1}{a^2} \,,
\end{equation}
which, combined with (\ref{ecc}) implies that
\begin{equation}\label{abe}
a = (1-e^2)^{-1/6}, \qquad b= (1-e^2)^{1/3}\,.
\end{equation}
The line element (\ref{ds'}) is then given by
\begin{equation}
    ds_{E}^2 = \frac{1}{(1-e^2)^{1/3}}\Big[(1-e^2\sin^2\chi)d\chi^2 + \sin^2\chi d\psi^2 \Big] \,,
\end{equation}
and (\ref{g}) becomes 
\begin{equation}
    h(\chi) = \int^{\chi}_{\pi/2}\sqrt{\csc^2v - e^2}\,dv\,.
\end{equation}

\subsection{Non-Uniform Flux Density Solutions}

We may now solve for the single particle LLL states on the surface of the squashed sphere. The resulting magnetic flux density $B(z,\bar{z})$ through the surface is non-uniform and depends on the eccentricity $e$ as well as the coordinates on the spheroid. This allows us to apply the framework outlined in Section 2.1.2 to solve the quantum problem. Before proceeding to do so, let's briefly consider the associated classical dynamics to build up some intuition.

\subsubsection{Classical Dynamics}

\begin{figure}[H]
    \begin{subfigure}{0.5\textwidth}
    \centering
	\includegraphics[scale=0.3]{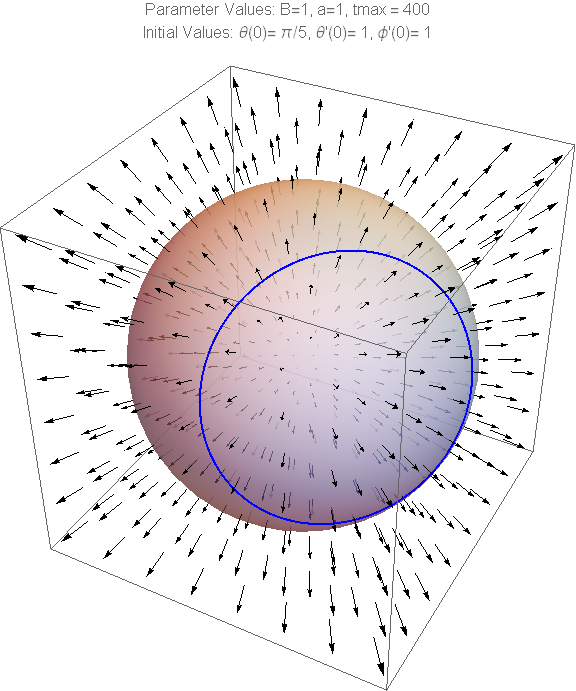} 
    \caption{$e=0$}\label{Dcl1} 
    \end{subfigure}
    \begin{subfigure}{0.5\textwidth}
    \centering
    \includegraphics[scale=0.3]{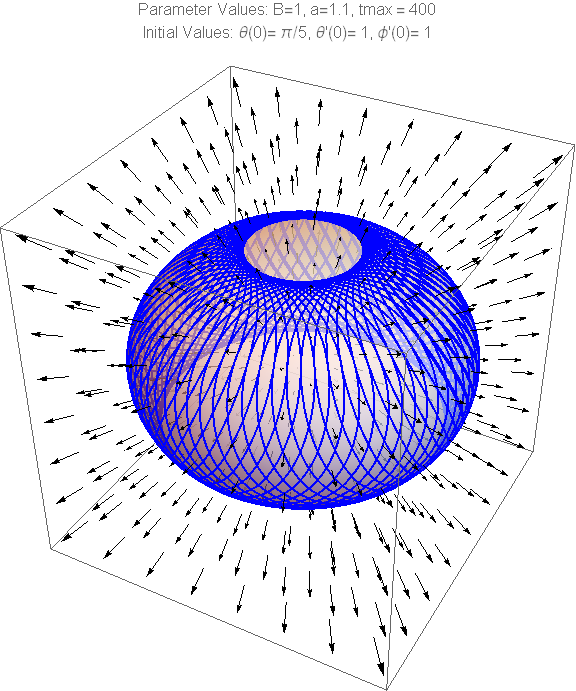} 
    \caption{$e=0.66$}\label{Dcl2}
    \end{subfigure}
	\\
    \begin{subfigure}{0.5\textwidth}
    \centering
   	\includegraphics[scale=0.3]{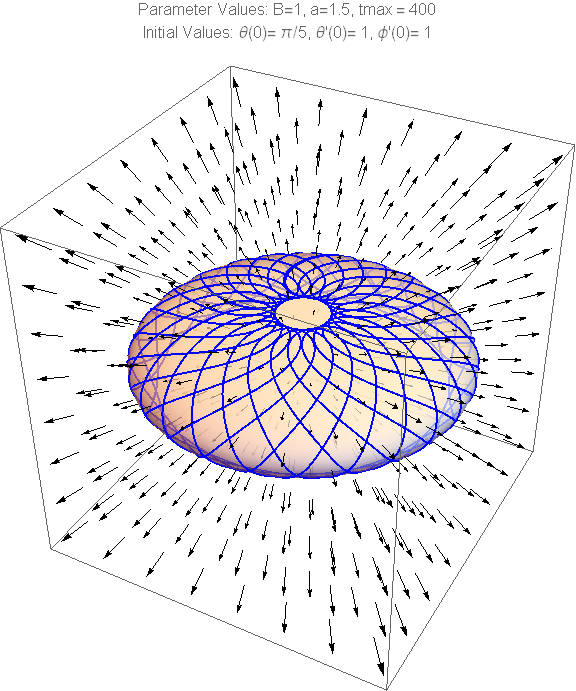} 
    \caption{$e=0.95$}\label{Dcl3}
    \end{subfigure}
    \begin{subfigure}{0.5\textwidth}
    \centering
   	\includegraphics[scale=0.3]{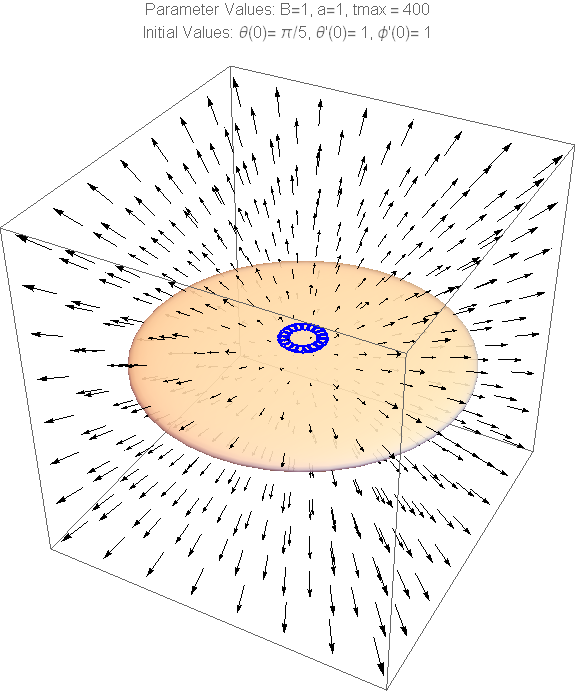} 
   \caption{$e=0.99$}\label{Dcl4}
    \end{subfigure}
    \caption{Classical trajectories for $B=1$ and various $e$.}\label{Dcl}
\end{figure}

The Lagrangian for a particle of charge $e$ and mass $m$ moving in a background magnetic field $\textbf{B} = \nabla \times \textbf{A}$ is given by
\begin{equation}
      L = \frac{1}{2}m \dot{\textbf{x}}^2 + e \dot{\textbf{x}} \cdot \textbf{A}\,.
\end{equation}
in units of $c=1$. The magnetic field associated to a monopole with field strength $B_0$ located at the origin is given by
\begin{eqnarray}\label{Bfield}
   \textbf{B} = \frac{B_0}{r^{2}}\widehat{\bm{r}}\,.
\end{eqnarray}
We choose to express the corresponding magnetic vector potential as
\begin{align}
    \textbf{A}(\theta,\phi) = \frac{B_0}{r}\frac{1-\cos\theta}{\sin\theta}(-\sin\phi,\cos\phi,0)\,.
\end{align}
The Dirac quantisation condition requires $B_0$ to be half-integer valued. With the particle constrained to move on the surface of the spheroid, the resulting equations of motion are given by
\begin{align}\label{eom}
0= &-256 e^2 (e^2-1) \bigg( e^2-1 - e^4 + 
     e^2 (e^2-2) \cos2\theta\bigg)\sqrt{
   1 - e^2 + \cot^2\theta} \sin2\theta \theta'^2 \nonumber\\
     & + 16 (2 - e^2 + e^2 \cos2\theta) \Bigg(\sqrt{2} B (1 - e^2)^{1/3}(2 - e^2 + e^2 \cos2\theta)^{7/2}
     \phi' \nonumber\\
   & - 4 (e^2-1) 
    (2 - e^2 + e^2 \cos2\theta) \sqrt{
    1 - e^2 + \cot^2\theta}
     \sin2\theta \phi'^2 \nonumber \\
     &- 8 (e^2-1) (-2 + 2 e^2 - e^4 + 
   e^2 (e^2-2) \cos2\theta) \sqrt{
    1 - e^2 + \cot^2\theta} \theta''\Bigg) \,,
    \\
0 = &2 (1 - e^2)^{1/6}
   \Bigg(B \sqrt{\frac{4 - 2 e^2 + 2 e^2 \cos2\theta}{
     1 - e^2 + \cot^2\theta)}} + \frac{
    8 (1 - e^2)^{2/3}
      \sin2\theta}{(2 - e^2 + e^2 \cos2\theta)^2} \phi' \Bigg) \theta' \nonumber\\
      &+ \frac{
 8 (1 - e^2)^{5/6}\sin^2\theta}{
 2 - e^2 + e^2 \cos2\theta} \phi'' 
     \,.
\end{align}
Unfortunately, we were unable to solve these equations for closed form analytic solutions.  We were however able to numerically solve them for a rich set of classical trajectories, a representative set of which are plotted in Figure \ref{Dcl}, with the magnetic monopole field illustrated in black and fixed at $B=1$. In the spherical limit where $e=0$, all classical trajectories are given by closed circular orbits whose radius, center and phase are all a function of initial conditions - illustrated for a particular set of initial conditions in Figure \ref{Dcl1}. We fix these initial conditions\footnote{Note while the initial conditions are fixed in terms of the initial values of $\phi$ and the polar angle $\theta$, they cannot be said to be equivalent since for different $e$ they correspond to points on different geometries.} and plot trajectories for various $e$ in Figure \ref{Dcl}. For slight squashing $e=0.66$, unlike in the spherical case the curved trajectory does not quite close into a circular orbit, resulting in the trajectory becoming highly delocalised over most of the spheroid. This delocalisation increases as $e$ is increased to $0.95$. However, for a ``severe" squashing where $e=0.99$, in which the surface of the spheroid  nears the limit of two parallel flat discs, at least for this set of initial conditions, the orbit becomes highly localised in a small band surrounding the pole. While different choices of initial conditions do not necessarily display localisation around the poles for large $e$, this particular behaviour will, in some sense, be mirrored in the quantum solutions that follow.

\subsubsection{Single Particle Quantum States}

We now proceed to solve the quantum problem. Recall that for general $g_s$, the term proportional to $B(z,\bar{z})$ in the Hamiltonian (\ref{ham}) breaks the degeneracy of the LLL. The solutions derived thus far will not be applicable. In the special case $g_s=2$ however, the LLL states $\psi$ must satisfy $\bar{\pi}\psi=0$, as before. Consequently, (\ref{Qstates}) furnishes a perfectly acceptable basis of LLL states. The flux density $B(z,\bar{z})$ of the magnetic monopole field through the deformed sphere, is given by the projection of the monopole field $\bm{B}$ onto the unit normal $\hat{\bm{n}}$ of the spheroid. Taking $|\bm{B}| = B_0/r^2$ and writing
\begin{align}
    \hat{\bm{n}} &= \frac{1}{\sqrt{b^2\sin^2\chi + a^2\cos^2\chi}}\big( b\sin\chi \cos\phi, b\sin\chi\sin\phi, a\cos\chi \big)\,,
    \end{align} 
yields
\begin{align}
  \quad B(\chi) &= \bm{B}\cdot\hat{\bm{n}} \nonumber    \\ 
    &= B_0 \frac{(1-e^2)^{1/3}}{(1-e^2\cos^2\chi)^{3/2}}\,,
\end{align}
written in terms of the eccentricity $e$. As may be expected, the magnetic flux of the monopole through the spheroid is azimuthal rotationally symmmetric, depending only on $\chi$. Note also that $B(\chi) \geq B_0$ at the poles, which move closer to the monopole as the sphere is squashed, while $B(\chi)\leq B_0$ at the equator, which is further away from the monopole than the corresponding point on the surface of the round sphere.
\\

\noindent
We can now solve the differential equation (\ref{Q})
\begin{equation}
    \frac{4}{\sqrt{g}}\partial\bar{\partial} Q(z,\bar{z}) = -\frac{2e}{\hbar}B(z,\bar{z})\,,
\end{equation}
for the magnetic potential $Q$. Substituting in for the metric (\ref{Qmetric}) and the magnetic field, the differential equation reads
\begin{equation}\label{main}
\frac{\partial}{\partial\chi}\Bigg( \Big[ \csc^2\chi-e^2 \Big]^{-1/2} \frac{\partial Q}{\partial\chi} \Bigg) = - \frac{2}{l^2} \sqrt{\frac{e^2-\csc^2\chi}{(e^2\cos^2\chi-1)^3}} \sin^2\chi \,,
\end{equation}
where we have defined an analogue of the magnetic length for this system $l=\sqrt{\hbar/eB_0}$. For a general value of the deformation parameter $e$, we were unable to find any closed form analytic solutions to this equation, so we resort instead to numerical integration. The integration constants are fixed by the requirement that the squashed sphere LLL states $\psi_m = z^me^{Q/2}$ reduce to those of the Haldane sphere in the $e\to 0$ limit. As an additional check, note that the Haldane state with $m=B_0$ is symmetrical about the equator of the sphere ($\theta=\pi/2$). Since the squashing of the sphere preserves the $\mathbb{Z}_2$ symmetry of the physical system with respect to the equator of the sphere, we expect the same symmetry to hold for the $m=B_0$ squashed sphere state for any value of $e$. We have verified numerically that this is indeed the case. Finally then, the squashed sphere LLL states are given by
\begin{align}\label{LLLQ}
    \psi_m(B_0,\epsilon;\theta,\phi) &= \mathcal{N} \exp(\frac{Q(\epsilon;\chi)}{2}+(B_0-m)h(\chi))e^{im\phi}\,, 
\end{align}
where $m = 0,1,2,...,2B_0$ and $\chi=\chi(\theta)$ as per (\ref{hire}). The normalization constant $\mathcal{N}$ is fixed by the requirement that $\bra{\psi_m}\ket{\psi_m}=1$ with respect to the inner product (\ref{prodQ}). Explicitly,
\begin{align}
    \langle\psi_n|\psi_m \rangle &= \int \psi_n \bar{\psi}_m dV \\
    &= 4R^2 \mathcal{N}^2 \int d\phi d\chi \frac{e^{-2h(\chi)}}{(1+e^{-2h(\chi)})^2} \sqrt{(b/a)^2 + \cot^2\chi}\, e^{Q/l^2-(n+m)h(\chi)}e^{i(n-m)\phi}\,,
\end{align}
using the volume element (\ref{volQ}) and the explicit form of the states (\ref{LLLQ}). Doing the $\phi$ integral, we obtain
\begin{align}
    \langle\psi_n|\psi_m \rangle &= 8\pi R^2 \mathcal{N}^2 \delta_{m,n} \int d\chi \frac{e^{-2(n+1)h(\chi)}}{(1+e^{-2h(\chi)})^2} \sqrt{(b/a)^2 + \cot^2\chi}\, e^{Q/l^2}\,,
\end{align}
Finally then, demanding that the states be orthonormal with respect to this inner product fixes
\begin{align}\label{Qnorm}
\mathcal{N}^{-2} &= 8\pi R^2 \int d\chi \frac{e^{-2(n+1)h(\chi)}}{(1+e^{-2h(\chi)})^2} \sqrt{(b/a)^2 + \cot^2\chi} e^{Q/l^2}\,.
\end{align}
Alas, this is another integral that we were unable to find a closed form expression for and $\mathcal{N}$ must be evaluated numerically on a case by case basis.

\subsection{Non-Uniform Flux Density Wavefunctions}

We can now plot various wavefunctions and perform a qualitative analysis of the effects of squashing the sphere on the basis of LLL states (\ref{LLLQ}). The fact that for different values of $e$, the states are defined over different geometries, complicates our task: plotting the states as functions of either $\chi$ or $\theta$ leads to distortions in the $e>0$ wavefunctions when represented in a two-dimensional Cartesian plot. These $e$-dependent distortions make comparison and hence analysis difficult. To circumvent this, we plot the probability distributions $|\psi|^2$ as a function of the \textit{normalized arclength} $\lambda(\theta)$, where $\lambda(0)=0$ and $\lambda(\pi)=1$. This arclength is equal to the (normalised) distance measured along a longitude on the surface of the ellipsoid. 
\\

\noindent
The normalisation (\ref{Qnorm}) is not appropriate for our plots as it distorts the distributions when plotted against $\lambda$. In the following we simply normalise the distributions such that
\begin{align}
   \int_0^1 |\psi|^2 d\lambda = 1
\end{align}
With this normalisation, the distributions represent the probability that the particle is located in the interval $\lambda+d\lambda$ at any fixed azimuth $\phi$. We consider various values of squashing parameter $e$, angular momentum $m$ and field strength $B_0= 1/l^2$ (in units of $e=\hbar=1$). Recall that the monopole field strength $B_0$ takes positive half-integer values, while the angular momentum is integer valued in the range $0\leq m \leq 2B_0$.\\

\begin{figure}
    \begin{subfigure}{0.5\textwidth}
    \centering
	 \includegraphics[scale=0.4]{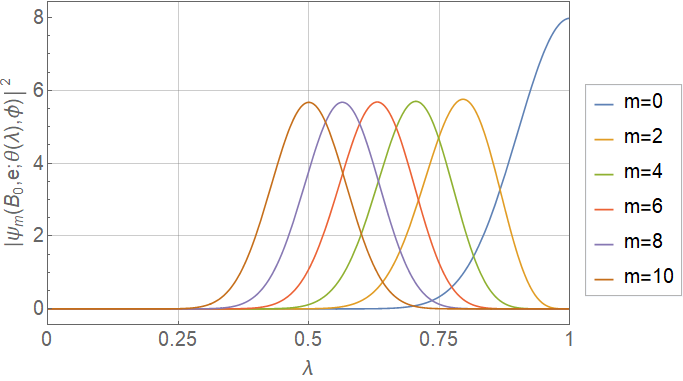} 
    \caption{$e=0$}\label{K1a}
   	\end{subfigure}
    \begin{subfigure}{0.5\textwidth}
    \centering
     \includegraphics[scale=0.4]{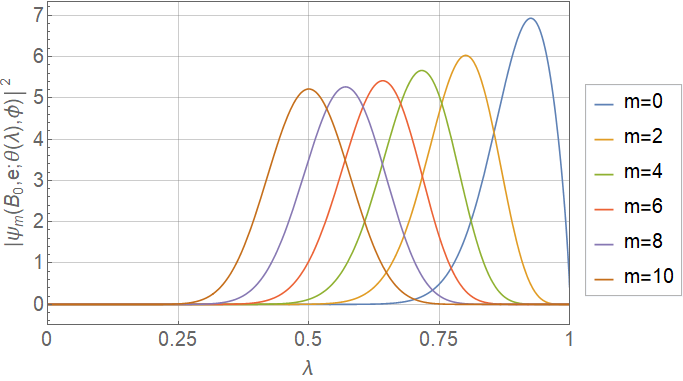} 
    \caption{$e=0.42$}\label{K1b}
    \end{subfigure}
       \begin{subfigure}{\textwidth}
    \centering
     \includegraphics[scale=0.4]{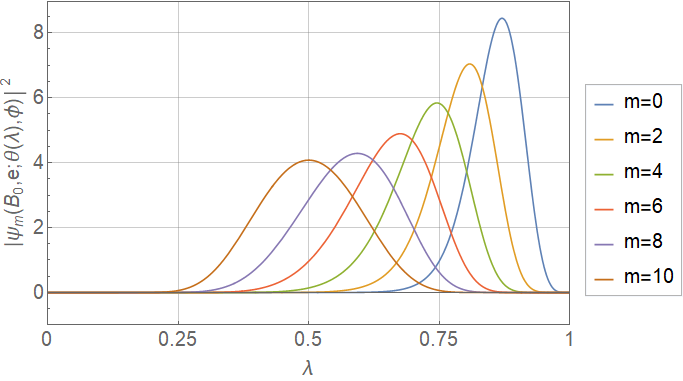} 
    \caption{$e=0.75$}\label{K1c}
    \end{subfigure}
    \begin{subfigure}{\textwidth}
    \centering
     \includegraphics[scale=0.4]{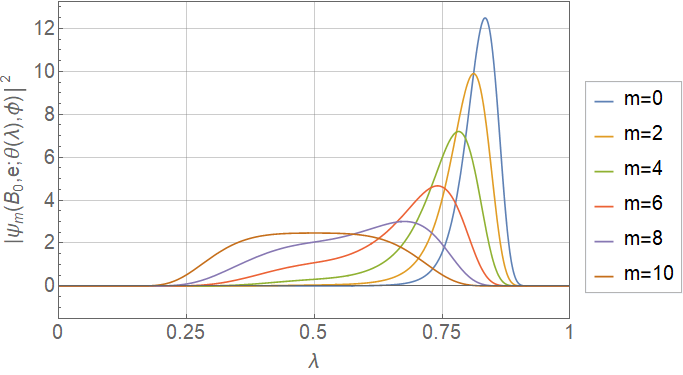} 
    \caption{$e=0.99$}\label{K1d}
    \end{subfigure}
    \caption{$|\psi_m(B_0,\epsilon;\lambda,\phi)|^2$ for $B_0=10$ and various $m$ and $e$.}\label{KK1}
\end{figure}

\noindent
We begin by fixing $B_0=10$ and plotting a representative set of states for various $m$ and $e$ in Figure \ref{KK1}. Note that, for general $e$, a state with quantum number $m=m'$ will always be equal to the state with quantum number $m=2B_0-m'$ upon reflection about the equator ($\lambda=0.5$) - this reflects the $\mathbb{Z}_2$ symmetry of the physical system with respect to reflection about the equator of the spheroid. In the following we therefore only plot states with $m\leq B_0$.
\\

\noindent
In the spherical limit, $e\to0$ we simply recover the usual Haldane states, which are evenly distributed over the surface of the sphere.\footnote{Note that the apparently larger amplitude of the $m=0$ state relative to the others is a result of our choice of coordinates, in which only half of the state is visible. On the sphere itself, the other half of this state lies on the other side of the pole, and the whole state is simply a translation by some amount $\theta$ of any of the $m\neq0$ states shown here. This agrees with our intuition regarding the $O(3)$ symmetry of the system, and the fact that the pole is indistinguishable from any other point on the sphere.} For $e=0.44$, the sphere has become squashed into an oblate spheroid, breaking the $O(3)$ rotational symmetry of the system. The poles, previously indistinguishable from any other point on the sphere, are now distinguished as the two points of minimum scalar curvature on the spheroid. Similarly, the equator now becomes the locus of maximum curvature. Inspecting the states we immediately notice two qualitative changes from the $e=0$ case: a) the amplitude of the $m=0$ state at the north pole ($\lambda=1$) has gone to zero, while its peak has shifted inward away from the pole; and b) the states with peaks closer to the poles have decreased in width and correspondingly increased in amplitude, with this localisation effect being more pronounced the closer the peak lies to the pole.
\\

\noindent
Increasing to $e=0.87$ (Figure \ref{K1c}) and $e=0.99$ (Figure \ref{K1d}) shows that the aforementioned effects become more pronounced with increasing squashing: the peak of the $m=0$ state moves significantly away from the pole and becomes highly localised, while the states closer to the equator delocalise over a larger fraction of the spheroid than before. Note how this phenomenon of localisation in a band centered on the pole was also recognised in our analysis of the classical trajectories.
\\

\begin{figure}
    \centering
     \includegraphics[scale=0.4]{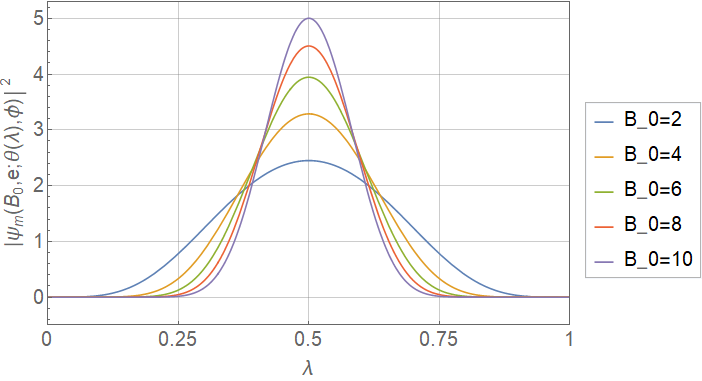}
    \caption{$|\psi_m(B_0,\epsilon;\lambda,\phi)|^2$ for $e=0.5$ and various $B_0=m$.}\label{KK1b}
\end{figure}

\noindent
In Figure \ref{KK1b} we fix $e=0.5$ and plot a set of $m=B_0$ states for various $B_0$, illustrating the tendency of states to become increasingly localised by stronger magnetic fields. This behaviour is characteristic of quantum states in constant magnetic fields, such as the planar QHE and the Haldane sphere.

\section{Summary}

We have applied the framework developed at the start of this article to find numerical results for the single-particle lowest Landau level states on the squashed sphere, parameterised by the spheroidal eccentricity $e$. The tendency of states to be increasingly localised by stronger magnetic fields (as in the spherical monopole system with Haldane's solutions) was observed to persist as the sphere is squashed. Additionally, we identified two qualitative features characterising the effect of squashing the sphere on the LLL states: the amplitude for finding a particle in the region immediately surrounding the poles goes to zero, while particles become increasingly localised in a band lying between the equator and the poles. \\

\noindent
Even after nearly a century of intense study, it seems particles in magnetic fields still have much to teach us about quantum states of matter. Many of these lessons have hinged on topological properties of the system and rightly so, given the remarkable progress made in topological quantum matter over the past two decades. However, as this work and its anachronistic sequel \cite{Murugan:2018hsd} have argued, there is also a rich geometric structure to the physics of the lowest Ladau level with implications not only in table-top condensed quantum matter but also perhaps in surface matter in astrophysical settings that realize extreme magnetic fields, such as neutron stars. More theoretically, following on from work on low-dimensional quantum gravity \cite{Maldacena:2018lmt}, Maldacena and collaborators constructed a new traversable wormhole solution in four spacetime dimensions \cite{Maldacena:2018gjk} in which the physics of Landau level states on a 2-sphere play a crucial role in stabilizing the black hole solution (leading to its traversability). Understanding how these states change under parametric deformations of the sphere (as we have studied here) would be of interest in determining the stability properties of the Maldacena-Milekhin-Popov wormholes. Clearly there is much still to be done, but we leave the exploration of these intriguing issues for future work.

\section{Acknowledgements}
JM is supported by the NRF of South Africa under grant CSUR 114599. RPS is supported by a graduate fellowship from the National Institute for Theoretical Physics.

\appendix

\section{Conformal Map from the Spheroid to the Plane}
In this appendix, we derive the map between spheroidal and planar coordinates used in the construction of wavefunctions on the deformed sphere. We start with the observation that the line element for any 2-dimensional surface can be put into the form 
\begin{equation}
   ds^2 = E(\theta,\phi)d\theta^2 + G(\theta,\phi)d\phi^2\,,
\end{equation}
with with an appropriate choice of orthogonal coordinates $(\theta,\phi)$ and where the functions $E$ and $G$ are known as the Gaussian fundamental functions, to be determined. For a round sphere of radius $R$, for example, the Gaussian fundamental functions are $e = R^{2}$ and $g = R^2\sin^2\theta$, giving the familiar line element 
\begin{align}
   ds^2 &= R^2 d\theta^2 + R^2\sin^2\theta d\varphi^2\,,
\end{align}
where $0<\theta<\pi$ and $0\leq\varphi<2\pi$. A spheroid, with equatorial and polar radii $a$ and $b$ respectively, has line element
\begin{align}
  ds'^2 &=(a^2\cos^2\chi + b^2\sin^2\chi)d\chi^2 + a^2\sin^2\chi d\phi^2\,, 
\end{align}
where $0<\chi<\pi$, $0\leq\phi<2\pi$, and Gaussian fundamental quantities
\begin{align}
E&= (a^2\cos^2\chi + b^2\sin^2\chi)\,, \quad G=a^2\sin^2\chi\,.
\end{align}
We would now like to construct a conformal map 
\begin{align}
\chi = \chi(\vartheta)\,, \quad \phi=\phi(\varphi)\,,
\end{align}
between these two surfaces. Demanding that this map be conformal imposes that the Gaussian fundamental quantities of the two metrics are related through \cite{26}
\begin{align}
\Big(\frac{\partial\vartheta}{\partial\chi}\Big)^2 \frac{E}{e} = \Big(\frac{\partial\varphi}{\partial\phi}\Big)^2 \frac{G}{g}
\end{align}
We first fix $\varphi=\phi$. Substitution then yields 
\begin{align}
\Big(\frac{\partial\vartheta}{\partial\chi}\Big)^2 \frac{R^2}{a^2\cos^2\chi + b^2\sin^2\chi}&=\frac{R^2\sin^2\vartheta}{a^2\sin^2\chi}\nonumber
\\
\Rightarrow \frac{d\vartheta}{\sin\vartheta}&= \frac{\sqrt{a^2\cos^2\chi + b^2\sin^2\chi}}{a\sin\chi}d\chi\,.
\end{align}
Integrating this expression yields
\begin{equation}\label{this}
\log\big(\tan\big(\frac{\vartheta}{2}\big)\big) - C = \int\sqrt{(b/a)^2 + \cot^2\chi}d\chi\,,
\end{equation}
where $C$ is an integration constant. We rewrite the right hand side as follows
\begin{align}
 \int\sqrt{(b/a)^2 + \cot^2\chi}d\chi = \int_{\pi/2}^{\chi}\sqrt{(b/a)^2 + \cot^2v}dv + \log(a/b)\,,
\end{align}
where the constant term is obtained by evaluating the anti-derivative of the integrand at $\chi = \pi/2$. Substituting back into (\ref{this}) yields
\begin{equation}
\log\Big(\tan\Big(\frac{\vartheta}{2}\Big)\Big)  = \int_{\pi/2}^{\chi}\sqrt{\Big(\frac{b}{a}\Big)^2 + \cot^2v}dv + \log(a/b) + C\,.
\end{equation}
Exponentiating and then inverting both sides, we obtain 
\begin{equation}
\cot\Big(\frac{\vartheta}{2}\Big) = A \frac{b}{a} e^{-h(\chi)}\,,
\end{equation}
where $A\equiv e^{-C}$, and we have defined
\begin{equation}
h(\chi) \equiv \int^{\chi}_{\pi/2}\sqrt{\Big(\frac{b}{a}\Big)^2 + \cot^2v}dv\,.
\end{equation} 
Imposing the initial condition $\vartheta(\pi/2)=\pi/2$ fixes $A=a/b$. Thus our conformal map from the sphere with coordinates $(\vartheta,\varphi)$ to the spheroid with coordinates $(\chi,\phi)$ is given by
\begin{equation} 
\cot\Big(\frac{\vartheta}{2}\Big) = e^{-h(\chi)}, \qquad \varphi=\phi\,.
\end{equation}
Finding the conformal map from the spheroid to the plane is now straightforward. The standard stereographic projection conformally maps the sphere of radius $R$ to the plane with standard polar coordinates $(r,\phi_p)$. This map is given by
\begin{equation}
r = 2R\cot\Big(\frac{\vartheta}{2}\Big)\,, \qquad \phi_p=\phi\,.
\end{equation}
To obtain a conformal mapping from the spheroid to the plane, we simply pull the stereographic projection back to the spheroid. The resulting conformal map is given by
\begin{equation}
r = 2R e^{-h(\chi)}\,, \quad \phi_p=\phi\,,
\end{equation}
where
\begin{equation}
h(\chi) = \int^{\chi}_{\pi/2}\sqrt{\Big(\frac{b}{a}\Big)^2 + \cot^2v}dv\,.
\end{equation}

\end{document}